\documentclass[a4paper]{article}

\usepackage{INTERSPEECH2020}
\usepackage{amsmath}
\usepackage{color, colortbl}
\usepackage{multirow}
\usepackage{graphicx}
\usepackage{subcaption}
\usepackage{hyperref}
\graphicspath{ {figures/} }

\newcommand{\G}{G}
\newcommand{\D}{D}
\newcommand{\Loss}{\mathcal{L}}
\newcommand{\K}{K} 
\newcommand{\Exp}{E} 
\newcommand{\xm}{x_m} 
\newcommand{\xa}{x_a} 
\newcommand{\ym}{y_m} 
\newcommand{\ymh}{\hat{y}_m} 
\newcommand{\xmh}{\hat{x}_m} 
\newcommand{\SDR}{SI-SDR}
\newcommand{\SDRi}{SI-SDRi}

\title{SEANet: A Multi-modal Speech Enhancement Network}
\name{Marco Tagliasacchi, Yunpeng Li, Karolis Misiunas, Dominik Roblek}

\address{
  Google Research
  }
\email{\{mtagliasacchi, yunpeng, kmisiunas, droblek\}@google.com}

\begin{document}

\maketitle
\begin{abstract}
We explore the possibility of leveraging accelerometer data to perform speech enhancement in very noisy conditions. Although it is possible to only partially reconstruct user's speech from the accelerometer, the latter provides a strong conditioning signal that is not influenced from noise sources in the environment. Based on this observation, we feed a multi-modal input to SEANet (Sound EnhAncement Network), a wave-to-wave fully convolutional model, which adopts a combination of feature losses and adversarial losses to reconstruct an enhanced version of user's speech. We trained our model with data collected by sensors mounted on an earbud and synthetically corrupted by adding different kinds of noise sources to the audio signal. Our experimental results demonstrate that it is possible to achieve very high quality results, even in the case of interfering speech at the same level of loudness. A sample of the output produced by our model is available at \mbox{\url{https://google-research.github.io/seanet/multimodal/speech}}. 
\end{abstract}

\section{Introduction}\label{sec:intro}
Enhancing the quality of speech is of paramount importance in digital communications. Speech degradation can occur for various reasons, e.g., from the interference of background noise, which can also contain overlapping speakers, to the effect of reverberations caused by room acoustics, to the artifacts introduced by compression and network impairments. This has motivated a very rich literature on speech enhancement and denoising.
Traditional signal processing methods adopt spectral noise subtraction~\cite{berouti1979, kamath2002}, spectral masking~\cite{reddy2007,grais2011},  statistical methods based on Wiener filtering~\cite{scalart1996} and Bayesian estimators~\cite{attias2001,loizou2005}. These methods make different assumptions about the underlying noise model (e.g., known signal-to-noise ratio (SNR), stationary noise, limited noise types, etc.), therefore they are unable to cope with challenging noisy conditions emerging when systems are deployed ``in-the-wild''. 

In recent years, data-driven methods have emerged, based on deep model architectures. Early works include methods based on denoising auto-encoders~\cite{feng2014} and recurrent models~\cite{weninger2015}. More recently, deep architectures have been adopted to improve speech enhancement based on spectral masking~\cite{wisdom2019}. Alternatively, generative models based on GANs~\cite{Pascual2017, donahue2017} and WaveNet~\cite{rethage2018} have been proposed. Speech denoising can also be seen as a special case of source separation, in which one of the sources represents the speech signal of interest~\cite{luo2017, kolbaek2017,leroux2019,kavalerov2019}. Our work belongs to the family of multi-modal models, which leverage additional conditioning signals to enhance the target speech. For example, the work in~\cite{hou2017} uses tight crops of mouth images to denoise speech. This approach was later extended by Looking2Listen~\cite{ephrat2018}, which uses visual information from facial crops segmented from videos to disentangle different speakers that talk simultaneously. A similar approach is presented in~\cite{ochiai2019}, which adopts an attention mechanism to weight the contribution of the audio and visual modalities. Multi-modal cues can also be exploited for voice activity detection~\cite{ariav2019}.

In this paper we consider the problem of multi-modal speech denoising. Instead of leveraging video as an additional modality, we consider data collected with a bone-conductance accelerometer mounted in an earbud, which operates synchronously with the microphone but at a lower sampling frequency. The sensor captures the local vibrations induced by the voice of the speaker, while being  relatively insensitive to external sources. Hence, it can be used as a conditioning signal to enhance user's speech and suppress noise. The fact that inertial measurement sensors mounted in mobile devices can be sensitive to speech has been recognized in the past literature. For example, gyroscope signals were used to recognize speech in~\cite{michalevsky2014}, while~\cite{anand2019} reconstructs speech from accelerometer-sensed reverberations induced by smartphone loudspeakers. The work in~\cite{hershey2004} combines signals from a microphone and a bone sensor using a Gaussian mixture model on the high-resolution log
spectra of each sensor. Similarly, multi-modal inputs are combined in~\cite{liu2018} using deep denoising autoencoders that reconstruct Mel-scale features fed to an ASR system. An ad-hoc speech recovery stage is needed to reconstruct the time-domain denoised waveforms. 

The proposed multi-modal SEANet (Speech EnhAncemnt Network) model receives two waveforms, one acquired with a microphone and one with an accelerometer, and produces as output a denoised speech waveform. The model is fully convolutional and maps waveforms to waveforms, without resorting to explicit time-frequency representations like short-time Fourier Transform (STFT) or mel spectrograms. To train the model, we adopt a combination of adversarial and reconstruction losses inspired by the recent MelGAN model~\cite{kumar2019}, which synthesizes waveforms from mel spectrograms. The adversarial losses induce the model to produce output waveforms that a discriminator cannot distinguish from clean speech. The reconstruction losses operate in the feature space defined by the discriminator and preserve speech content while suppressing noise. 

In our experiments we consider challenging scenarios in which the target speech signal is mixed with that of other speakers, or different kinds of background noise sampled from Freesound~\cite{font2013}. We demonstrate that by leveraging the conditioning signal collected by the accelerometer, it is possible to denoise speech even in very adverse conditions. We collected a dataset that contains speech and the corresponding accelerometer readings and observed an improvement in scale-invariant signal-to-distortion ratio (\SDRi) of 9.6dB when the interferer is mixed with a unit gain. 
\begin{figure*}[t]
\centering
\includegraphics[width=0.9\textwidth]{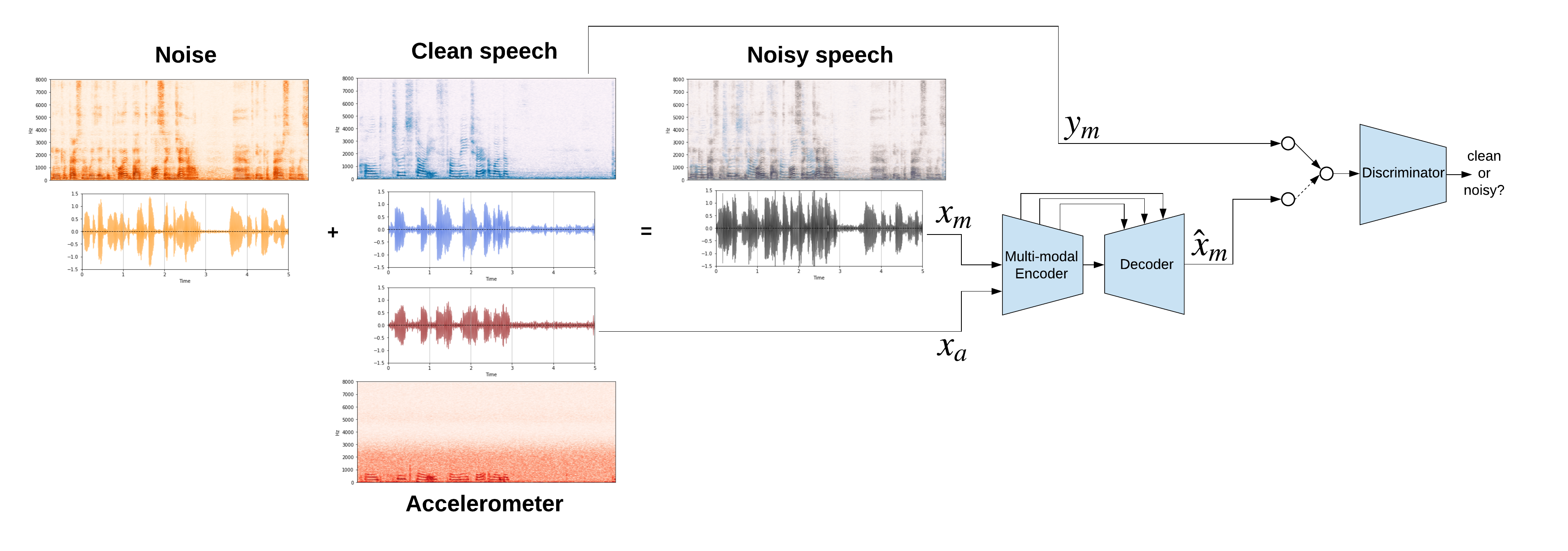}
\caption{SEANet model overview. A noisy speech signal, obtained superimposing clean speech with a noise source, is fed to the multi-modal encoder together with the accelerometer signal. Spectrograms are shown only for illustration purposes, as they are not explicitly computed by the proposed wave-to-wave model. }\label{fig:model_overview}
\end{figure*}

\section{Method}
\label{sec:method}
The proposed SEANet model is trained in a fully supervised fashion using pairs $\langle(\xm, \xa), \ym\rangle$, where $\xm$ denotes the input noisy speech collected by the microphone, $\xa$ the accelerometer signal used as conditioning, and $\ym$ the target audio signal corresponding to clean speech. Note that $\xa$ might have one or more channels, depending on the number of accelerometer axes used. We assume that $\xm$, $\xa$ and $\ym$ are time-aligned and available at the same sampling rate. Since the sampling rate of accelerometers is typically smaller than that of the microphone, the former signal is interpolated before being fed to the model.

The model architecture consist of a UNet generator $\G(\xm, \xa)$, which take as its input an audio $\xm$ and one or more accelerometer readings $\xa$ time-aligned with the audio.
In Figure~\ref{fig:model_overview} we illustrate the case in which a single accelerometer axis is used. The generator produces as output a single-channel waveform $\xmh$, which represents the denoised speech. The discriminator is asked to determine whether its input comes from the distribution of clean speech, or from the output of the generator. 

\begin{figure}[t]
\centering
\includegraphics[width=0.405\textwidth]{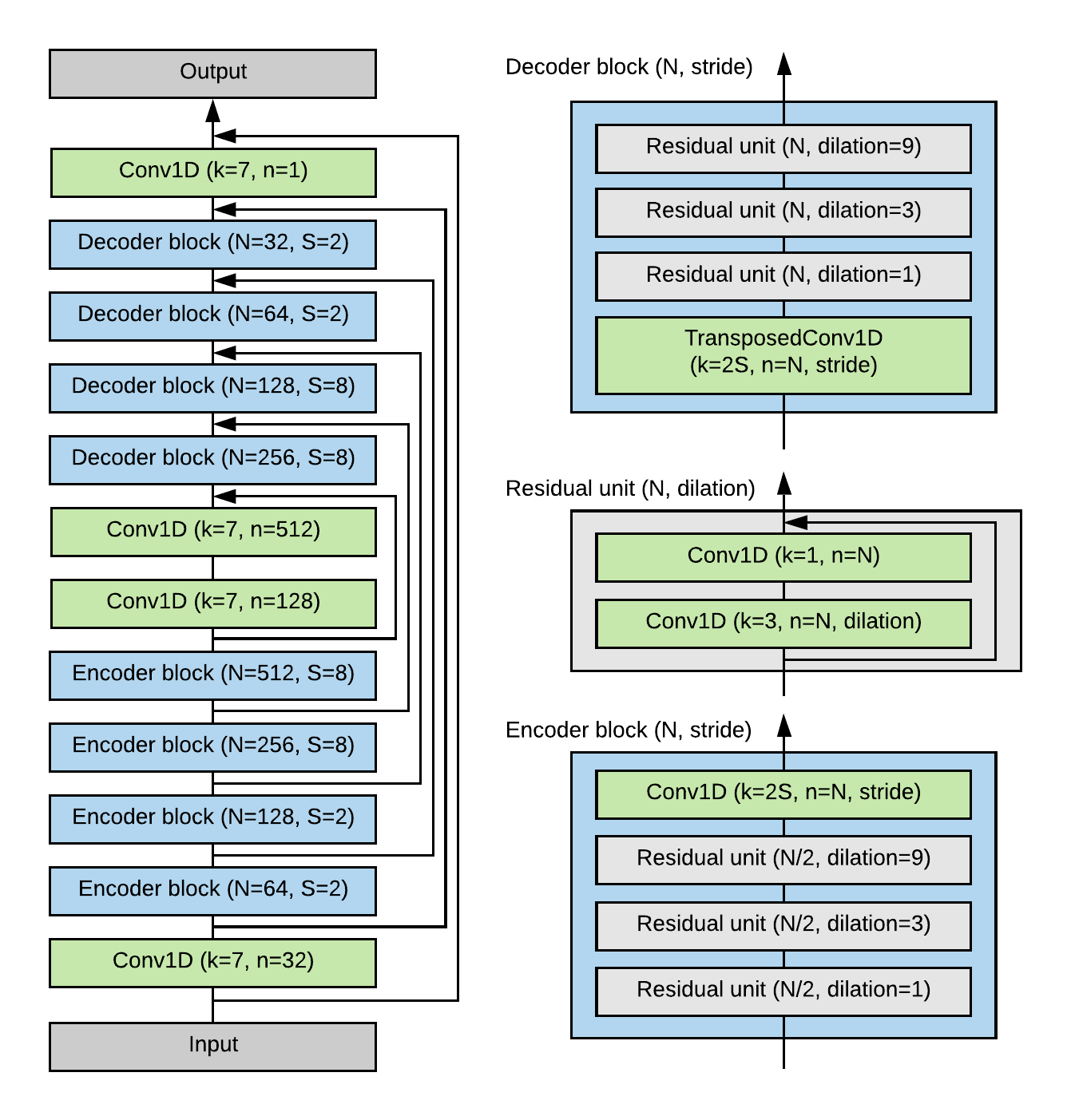}
\caption{Generator architecture.}
\label{fig:generator_architecture}
\end{figure}

\textbf{Model architecture}:~Our UNet generator is a symmetric encoder-decoder network with skip-connections. The decoder adopts the same architecture as the generator in~\cite{kumar2019}, while the encoder  mirrors the decoder in its layout. A skip-connection is added between each encoder block and its mirrored decoder block. The out-most skip connects only the speech channel needed by the output.
The encoder and the decoder have each four blocks stacked together, which are sandwiched between two plain convolution layers. The encoder follows a down-sampling scheme of (2, 2, 8, 8) while the decoder up-samples in the reverse order. The number of channels is doubled whenever down-sampling and halved whenever up-sampling.
Each decoder block consists of an up-sampling layer, in the form of a transposed 1D convolution, followed by three residual units each containing 1D convolutions with dilation rates of 1, 3, and 9, respectively. The encoder block again mirrors the decoder block, and consists of the same residual units followed by a strided 1D convolution for down-sampling. 
The overall structure of the generator is illustrated in Figure~\ref{fig:generator_architecture}.

For the discriminator, we use the same multi-resolution convolutional architecture as~\cite{kumar2019}.
Three structurally identical discriminators are applied to input audio at different resolutions: original, 2x down-sampled, and 4x down-sampled.
Each discriminator consists an initial plain convolution followed by four grouped convolutions~\cite{krizhevsky2012}, each of which has a group size of 4, a down-sampling factor of 4, and a channel multiplier of 4 up to a maximum of 1024 output channels. They are followed by two more plain convolution layers to produce the final output, i.e., the logits. Note that since the discriminator is fully convolutional, the number of logits in the output is more than one and proportional to the length of the input audio. Each logit judges the plausibility of a segment of the input that corresponds to its receptive field.
We refer interested readers to~\cite{kumar2019} for more architectural details.

We use weight normalization~\cite{salimans2016} and ELU activation~\cite{clevert2016} in the generator, while layer normalization and Leaky ReLU activation~\cite{maas2013} with $\alpha=0.3$ are used in the discriminator.

\textbf{Loss functions}:~SEANet combines adversarial and reconstruction losses to train simultaneously the generator and the discriminators. 
The adversarial loss is a hinge loss averaged over multiple resolutions and over time.
More formally, let $k\in\{1,\dots,K\}$ index over the individual discriminators for different resolutions, and $t$ index over the length of the output, i.e., the number of logits $T_k$, of discriminator $k$. The discriminator loss can be written as
\begin{align}
    \nonumber
    \Loss_\D =~&\Exp_{\ym} \left [\frac{1}{\K}\sum_{k,t} \frac{1}{T_k} \max(0, 1 - \D_{k,t}(\ym)) \right ] + \\  
    &\Exp_{(\xm, \xa)} \left [ \frac{1}{\K}\sum_{k,t} \frac{1}{T_k} \max(0, 1 + \D_{k,t}(G(\xm, \xa)) \right ],  
\end{align}
while the adversarial loss for the generator is
\begin{equation}
    \Loss_G^{\text{adv}} = \Exp_{(\xm, \xa)} \left [ \frac{1}{\K}\sum_{k,t} \frac{1}{T_k} \max(0, 1 - \D_{k,t}(G(\xm, \xa)) \right ].  
\end{equation}

For the reconstruction loss we use the ``feature'' loss proposed in~\cite{kumar2019}, namely the normalized L1 distance between the discriminator internal layer output of the generator audio and that of the corresponding target audio:
\begin{equation}
    \Loss_G^{\text{rec}} = \Exp_{x} \left [ \frac{1}{\K}\sum_{k,l} \frac{1}{L} \frac{\| \D_k^{(l)}(\ym) - \D_k^{(l)}(\G(\xm, \xa)) \|_1}{T_{k,l}} \right ],
\end{equation}
where $x \triangleq \langle(\xm, \xa), \ym\rangle$ denotes a training example, $L$ is the number of internal layers, $D_k^{(l)}$ for $l\in\{1,\dots,L\}$ is the output of layer $l$ of discriminator $k$, and $T_{k,l}$ is length of the feature layer $D_k^{(l)}$.
Compared with per-sample losses, such as the average L1 distance between waveforms, the feature loss tends to be less sensitive to small misalignment.
The overall generator loss is a weighted sum of the adversarial and the reconstruction loss, i.e.,
\begin{equation}
    \Loss_\G = \Loss_G^{\text{adv}} + \lambda \cdot \Loss_G^{\text{rec}}.
\end{equation}

For all our experiments, we set the weight of the reconstruction loss $\lambda$ to $100$ and use a discriminator with $K =3$ scales. We train with the Adam optimizer, with a batch size of 16 and a constant learning rate of 0.0001 with $\beta_1=0.5$ and $\beta_2=0.9$. We train for 200k iterations (2M iterations when training on Librispeech) on a single GPU. We evaluate results using the last checkpoint of each training run. No parameter tuning or early stopping were performed.

\section{Experiments}\label{sec:experiments}
\textbf{Datasets}:~We collected an in-house dataset with sensors mounted on an earbud, since a dataset with these characteristics is not available in the literature. The microphone sampled audio waveforms at 16kHz, while the 2-axis accelerometer operated at 4kHz. We selected one of the two axes and interpolated the accelerometer signal at 16kHz before feeding it to the model. We then applied high-pass filtering with a cut-off of 20Hz to all signals and normalized the amplitudes dividing all samples by a factor $1.1 \cdot \text{quantile}(x; 0.9999)$ and clipping the result in the $[-1, +1]$ range. This is necessary to deal with isolated spikes which were present in the raw output of the accelerometer. 

\begin{figure}
\centering
\includegraphics[width=0.45\textwidth]{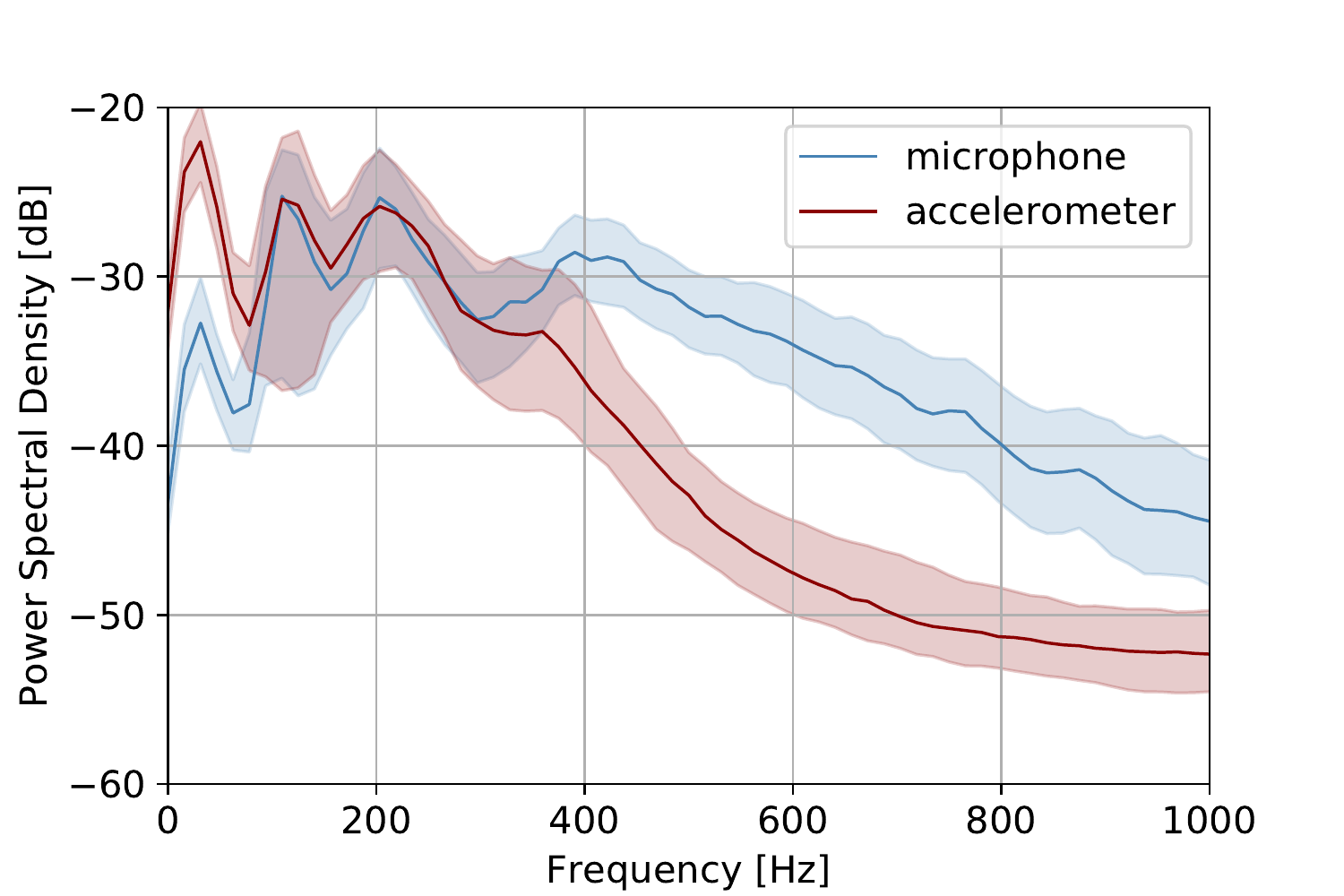}
\caption{Power spectral density: microphone vs. accelerometer.}
\label{fig:power_spectral_density}
\end{figure}

We asked 25 subjects to speak while wearing one earbud in a relatively quiet office environment. In total we collected $\sim$1.25 hours of data, with each subject contributing $\sim$3 minutes. We organized the data in 5 folds, so that in each fold 20 speakers are used for training and 5 speakers for testing. 
Figure~\ref{fig:power_spectral_density} shows the power spectral density of the signals acquired by the sensors. We observe that they share a similar response in the range of 100--300Hz, while the sensitivity of the accelerometer decreases rapidly above 300Hz. 

To explore the quality potentially achievable if we had access to more data, we created a synthetically generated multi-modal dataset.  First, we trained a variant of SEANet which learns to map audio waveforms to the corresponding accelerometer waveform, using the in-house dataset described above. This model uses the same architecture described in Section~\ref{sec:method}, with the only difference that it receives one input channel with clean audio and produces one output channel with the corresponding accelerometer signal. Note that learning this mapping is much easier than reconstructing audio samples from the accelerometer signal alone. Then, we fed audio samples from Librispeech~\cite{panayotov2015} to synthesize the corresponding accelerometer signal. In this case, we followed the canonical split provided by Librispeech, using \emph{train-clean-100} for training and \emph{test-clean} for testing.

To generate the noisy input $\xm$, we mix the clean microphone recording $\ym$ with other noise sources. We consider two scenarios: i) \emph{mixed speech}, in which an utterance from a different speaker is mixed with the clean source; ii) \emph{mixed noise}, in which we mix with samples taken at random from Freesound~\cite{font2013}, to mimic a wide and diverse range of noise sources, with a unit mixing gain. 
In one of the experiments, we also limit the bandwidth of the accelerometer to simulate a sensor operating at lower sampling rates. In this case we use the following downsampling factors \{16, 20, 32, 40, 50, 64, 80, 100\}, corresponding to the sampling frequencies \{1000Hz, 800Hz, 500Hz, 400Hz, 320Hz, 250Hz, 200Hz, 160Hz\}. We also report results of an audio-only SEANet model, in which the accelerometer input is not used.

\textbf{Metrics and baselines}:~In order to evaluate the quality of the enhanced speech, we measure the scale invariant signal-to-distortion ratio (\SDR), which accommodates for an amplitude gain mismatch between the estimated signal $\ymh$ and the ground truth clean reference signal $\xm$. The \SDR~is computed as described in~\cite{kavalerov2019}.

We evaluated models recently proposed in the speech enhancement and separation literature, which receive as input only the audio signal. It is worth noting that a direct comparison with these methods is not meaningful, as SEANet receives as input an additional conditioning signal. However, this evaluation is useful to gauge the level of complexity of the dataset, highlighting the added value of leveraging the accelerometer signal. Namely, we include in our evaluation iTDCN++~\cite{kavalerov2019} and Wavesplit~\cite{zeghidour2020}. The iTDCN++ model is inspired by Conv-TasNet and predicts a mask with a sigmoid activation that is applied to the mixture STFT coefficients. Wavesplit infers and clusters representations of each speaker and then estimates each source signal conditioned on the inferred representations.

\begin{table}[t]
    \centering
    {\footnotesize
    \caption{Mean \SDRi~for the In-house dataset.}
    \label{tab:SDR_splits}
    \begin{tabular}{lcccc}
    \hline
    scenario & split & SEANet & SEANet \\ 
     & & audio + accel & audio only \\ 
    \hline
\multirow{5}{*}{Mixed noise} 
 & 1 & $9.9 \pm 0.2$ & $8.4 \pm 0.2$ \\
 & 2 & $8.0 \pm 0.2$ & $7.9 \pm 0.1$ \\
 & 3 & $8.3 \pm 0.1$ & $7.2 \pm 0.2$ \\
 & 4 & $8.8 \pm 0.1$ & $8.1 \pm 0.1$ \\
 & 5 & $9.9 \pm 0.1$ & $8.4 \pm 0.1$ \\
& avg. & $8.9$ & $8.0$ \\
\hline
\multirow{5}{*}{Mixed speech} 
& 1 & $10.1 \pm 0.1$ & $-0.9 \pm 0.1$ \\
 & 2 & $8.6 \pm 0.1$ & $-0.9 \pm 0.1$ \\
 & 3 & $9.2 \pm 0.1$ & $-0.7 \pm 0.0$ \\
 & 4 & $9.0 \pm 0.2$ & $-1.0 \pm 0.1$ \\
 & 5 & $11.1 \pm 0.2$ & $-0.9 \pm 0.1$ \\
 & avg. & $9.6$ & $-0.9$ \\
    \hline
\end{tabular}
}

\end{table}

\begin{table}[t]
    \centering
{\footnotesize
    \caption{Mean \SDRi~for Librispeech. }
    \label{tab:SDR_librispeech}
    \begin{tabular}{lcccc}
        \hline
        scenario & split & SEANet & SEANet \\ 
         & & audio + accel & audio only \\ 
        \hline
        Mixed noise
         & test & $12.4 \pm 0.3$ & $9.8 \pm 0.2$ \\
        Mixed speech
        & test & $12.4 \pm 0.3$ & $-1.0 \pm 0.2$ \\
        \hline
    \end{tabular}
}
\end{table}

\begin{figure}[t]
    \centering
    \begin{subfigure}[b]{0.39\textwidth}
        \includegraphics[width=1.0\textwidth]{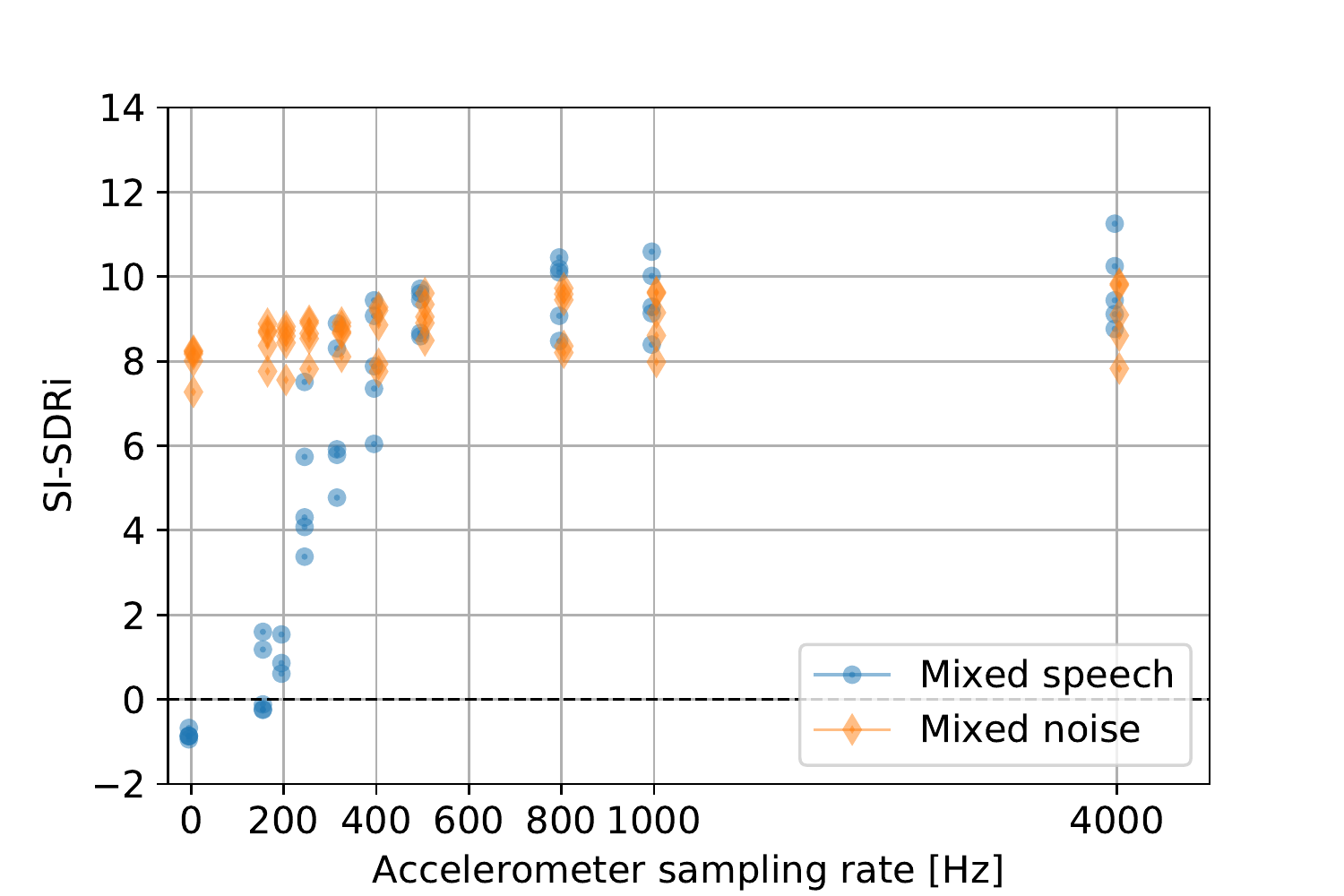}
        \caption{In-house dataset.}
        \label{fig:SDR_cutoff}
    \end{subfigure}
    ~ 
    \begin{subfigure}[b]{0.39\textwidth}
\includegraphics[width=1.0\textwidth]{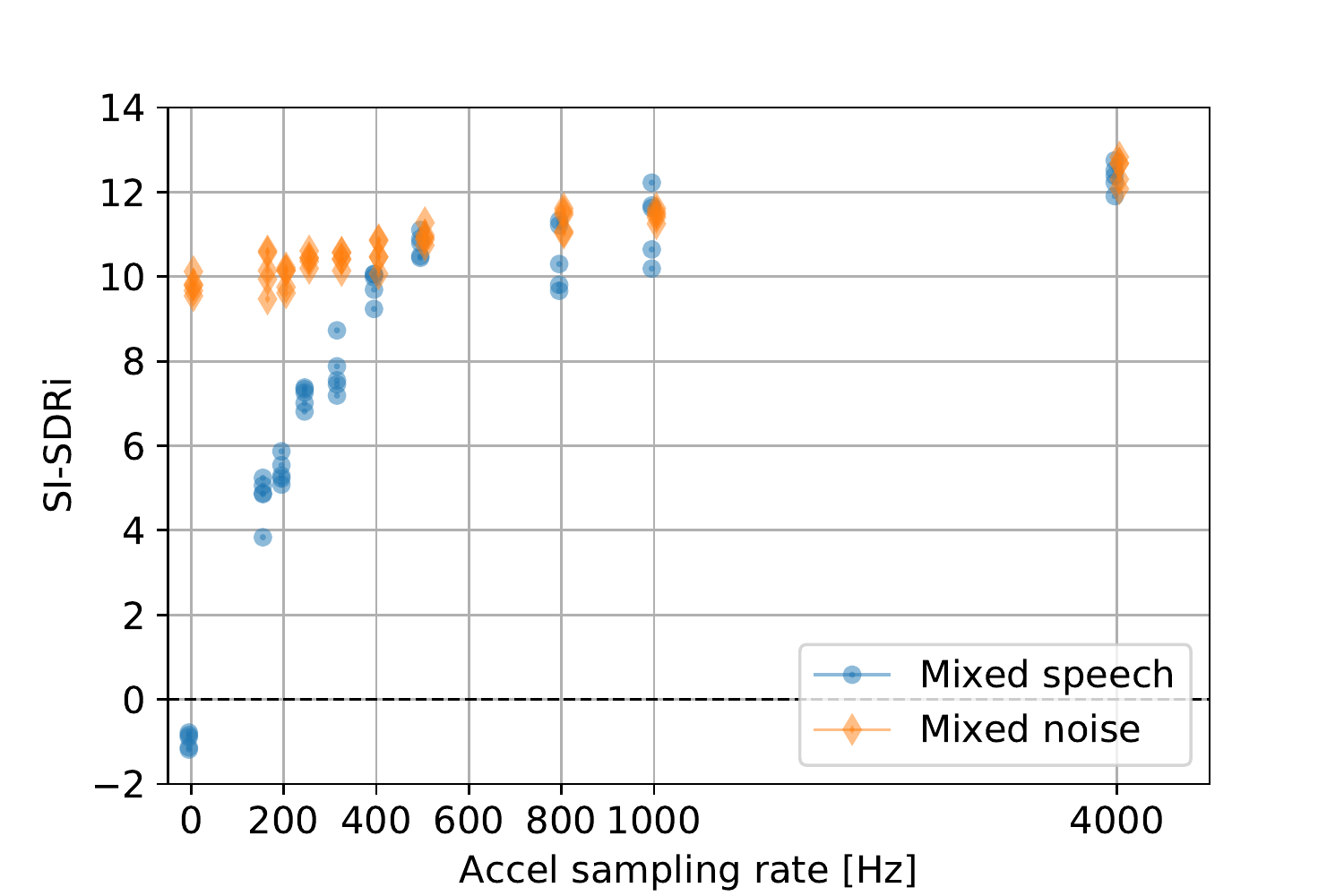}
\caption{Librispeech.}
\label{fig:SDR_synth_cutoff}
    \end{subfigure}
    \caption{Improvement in \SDR~for different accelerometer sampling rates (each point represents one replica).}\label{fig:SDR_cutoff_all}
\end{figure}

\textbf{Results}:~Table~\ref{tab:SDR_splits} reports the results obtained repeating five replicas, on each of the five splits for the two scenarios. The average \SDRi~is 8.9dB when mixing with background noise from Freesound and 9.6dB when mixing with speech. Note that the variability across replicas is small (standard deviation $\pm 0.1-0.2$dB), while there is a more significant variability across splits. We repeated the experiment by changing the gain used during mixing and observed that the \SDRi~varies between 3.7dB (6.2dB), at 10dB mixing gain, and 15.0dB (15.1dB), at -10dB mixing gain for mixing noise (mixing speech). 
Table~\ref{tab:SDR_splits} includes results when SEANet is trained using audio only. In the mixed noise scenario, the model is still able to enhance speech, although attaining a lower \SDRi~(7.9dB vs. 8.9dB). Conversely, in the mixed speech scenario the audio-only variant of SEANet is unable to separate the speakers. This is not surprising, since the model as described in this paper does not include a permutation invariant loss, which is needed to separate sources of the same kind. Using audio-only, iTDCN++ attains 7.5dB on mixed noise (trained on synthetically reverberated Libri-Light speech + synthetically-reverberated Freesound) and 4.2dB on mixed speech (trained on synthetically reverberated Libri-Light speech mixtures), while Wavesplit attains 8.8dB on mixed speech (trained on Librispeech mixtures, with no reverberation). This demonstrates the inherent difficulty of the in-house dataset and the fact that the availability of the conditioning signal makes the denoising problem significantly easier, especially in the scenario with mixed speech. 

We also evaluate a model trained on Librispeech with synthetically generated accelerometer signals. Table~\ref{tab:SDR_librispeech} shows that this model achieves an \SDRi~of 12.4dB on both mixed noise and mixed speech, thus hinting to the fact that better accuracy can be attained using a larger dataset during training. Examples of the denoised results produced by SEANet are publicly available at the following page: \mbox{\url{https://github.com/google-research/seanet/multimodal/speech}}.

We investigated the contribution of the conditioning provided by the accelerometer. To this end, we progressively decimated the accelerometer signal before feeding it to our model during both training and evaluation. Figure~\ref{fig:SDR_cutoff} shows an interesting result. In the scenario with two overlapping speakers, a rapid decrease in \SDRi~is observed when the sampling rate drops below 400Hz, and our model is unable to separate the speakers when the sampling rate is smaller than 200Hz. Conversely, for the scenario with background noise, only a small decrease in \SDRi~is observed, also when the sampling rate of the accelerometer is drastically reduced. The average \SDRi~across the splits drops from 8.9dB to 8.0dB. We can argue that this is a simpler scenario, giving the distinct acoustic characteristics of the background noise. These results are confirmed when training and evaluating on the multi-modal dataset generated from Librispeech, as illustrated in Figure~\ref{fig:SDR_synth_cutoff}. In this case the 
average \SDRi~drops from 12.4dB to 9.8dB.
\section{Conclusions}\label{sec:conclusions}
In this paper we show that the accelerometer data collected from sensors mounted on earbuds provides a strong conditioning signal for speech denoising. This is especially useful in the challenging scenario with overlapping speakers. In our future work we plan to expand the multi-modal aspect of SEANet by exploring how to combine multiple microphone signals, accelerometer axes and visual cues.

\section{Acknowledgements}
We would like to thank Kevin Wilson, Scott Wisdom, John Hershey, Dick Lyon and Neil Zeghidour for their help with and feedback on this work. We also thank Alina Mihaela Stan for the help collecting the in-house dataset.

\bibliographystyle{IEEEtran}

\bibliography{references}

\end{document}